# Detecting Dark Patterns in User Interfaces Using Logistic Regression and Bag-of-Words Representation

*Aliyu Umar[1], Maaruf Lawan[2], Adamu Lawan[3], Mukhtar Dahiru[2], Abdullahi Abdulkadir[4]

[1] University of Portsmouth, Portsmouth, PO1 2UP UK
[2] Jigawa State Polytechnic for ICT. Kazaure, Jigawa, Nigeria
[3] Beihang University
[4] Shehu Shagari College of Education Sokoto

Correspondence should be addressed to Aliyu Umar;

aliyu.umar@port.ac.uk*

**Abstract.** Dark patterns in user interfaces represent deceptive design practices intended to manipulate users' behavior, often leading to unintended consequences such as coerced purchases, involuntary data disclosures, or user frustration. Detecting and mitigating these dark patterns is crucial for promoting transparency, trust, and ethical design practices in digital environments. This paper proposes a novel approach for detecting dark patterns in user interfaces using logistic regression and bag-of-words representation. Our methodology involves collecting a diverse dataset of user interface text samples, preprocessing the data, extracting text features using the bag-of-words representation, training a logistic regression model, and evaluating its performance using various metrics such as accuracy, precision, recall, F1-score, and the area under the ROC curve (AUC). Experimental results demonstrate the effectiveness of the proposed approach in accurately identifying instances of dark patterns, with high predictive performance and robustness to variations in dataset composition and model parameters. The insights gained from this study contribute to the growing body of knowledge on dark patterns detection and classification, offering practical implications for designers, developers, and policymakers in promoting ethical design practices and protecting user rights in digital environments.

## 1 Introduction

In the digital age, User Interfaces (UIs) play a pivotal role in shaping user experiences across various online platforms, ranging from e-commerce websites to social media platforms and mobile applications. However, amidst the pursuit of enhancing user engagement and conversion rates, some designers and developers resort to employing deceptive UI design techniques known as "dark patterns" [1], [2]. Dark patterns are user interface design choices crafted to manipulate or coerce users into taking actions they might not otherwise choose to do [3]. These patterns exploit cognitive biases and psychological vulnerabilities to nudge users towards decisions that serve the interests of the interface's designers, often at the expense of user autonomy, privacy, and well-being [4].

The proliferation of dark patterns poses significant challenges to user trust, privacy, and digital well-being. From misleading advertisements and hidden costs to forced opt-ins and deceptive interfaces, the prevalence of dark patterns undermines user trust in online platforms and erodes the fundamental principles of transparency and user empowerment [5]. Moreover, dark patterns can have tangible consequences beyond mere annoyance, leading to unintended purchases, privacy violations, and negative user experiences [6].

Detecting and mitigating dark patterns is thus imperative to safeguarding user rights and fostering a fair and transparent digital environment. While manual detection and classification of dark patterns are labor-intensive and subjective, recent advancements in machine learning and natural language processing offer promising avenues for automated detection of deceptive UI practices [7]. By leveraging computational techniques, such as logistic regression and bag-of-words representation, researchers aim to develop robust and scalable models capable of identifying and flagging instances of dark patterns in digital interfaces [7].

This paper presents a novel approach to detecting dark patterns in user interfaces using logistic regression and bag-of-words representation. Through a comprehensive analysis of textual UI elements and machine learning techniques, we demonstrate the efficacy of our model in identifying deceptive design patterns and empowering users to make informed decisions online. By shedding light on the prevalence of dark patterns and offering a data-driven solution for their detection, this research contributes to enhancing user trust, privacy, and digital well-being in the ever-evolving landscape of digital interfaces.

## 2  Related Work:

The detection and classification of dark patterns in user interfaces have garnered significant attention from researchers and practitioners in the fields of human-computer interaction (HCI), computer science, and digital ethics. This section provides an overview of existing literature on dark patterns detection and classification methods, highlighting key contributions, methodologies, and insights [8].

Early research efforts focused on conceptualizing and defining the notion of dark patterns, shedding light on the deceptive design practices employed by online platforms to manipulate user behavior [1]. Brignull (2013) [9] introduced the term "dark patterns" and categorized deceptive UI design techniques into various typologies, including misdirection, forced action, and sneak into basket [2]. Subsequent studies by [5] and [10]) further formalized the concept of dark patterns and explored its implications for user privacy and digital well-being.

In recent years, researchers have increasingly turned to machine learning techniques to automate the detection and classification of dark patterns in digital interfaces. In[11] a machine learning-based approach for identifying dark patterns in e-commerce websites is proposed, leveraging features extracted from UI elements and user interactions [12]. Their study demonstrated the feasibility of using supervised learning algorithms, such as logistic regression and random forest, to distinguish between deceptive and non-deceptive UI patterns with high accuracy.

Despite these advancements, challenges remain in effectively detecting and mitigating dark patterns, particularly in the context of evolving UI design trends and user preferences. Future research directions may include exploring novel data-driven approaches, such as deep learning and adversarial machine learning, for more robust and adaptive detection of deceptive UI patterns [8]. Additionally, interdisciplinary collaborations between HCI researchers, data scientists, and ethicists are essential for addressing the ethical implications and societal impact of dark patterns on user trust and digital well-being.

Logistic Regression, a foundational classification technique, has been utilized in early studies to detect dark patterns. While its simplicity allows for interpretability, its performance is often benchmarked against more complex models. In comparison to methods like Random Forests or Transformers, Logistic Regression typically demonstrates moderate performance, achieving accuracies around 91% in controlled datasets of deceptive patterns in e-commerce [11]. Explainability in detecting dark patterns is a growing concern. Methods such as SHAP (SHapley Additive exPlanations) and LIME (Local Interpretable Model-agnostic Explanations) are integrated into detection frameworks to improve transparency. For example, recent studies applied these techniques to Logistic Regression and other classifiers to explain how features like misleading text or visual emphasis contribute to the detection of deceptive elements [13], [14].

Publicly available datasets, such as the E-Commerce Dark Pattern Dataset, have been instrumental in advancing research. These datasets include labeled instances of dark and non-dark patterns, enabling robust model training and evaluation. The dataset facilitates comparative studies, showing that Logistic Regression with BoW can serve as a reliable baseline, although deep-learning models like RoBERTa often exhibit superior performance in larger datasets [7], [15]

## 3   Methodology:

### 3.1   Dataset Collection

The dataset was obtained from GitHub repository[1]. It is a text-based dataset for automatic dark pattern detection, originally from [1] study, comprising 3,636 text instances, evenly split between dark patterns and non-dark patterns. A diverse set of user interface text samples was collected from various online platforms, including e-commerce websites, social media platforms, and mobile applications. The dataset aimed to encompass a wide range of UI design patterns, including both benign and deceptive practices. Each text instance was manually labeled by human annotators as either a dark pattern or a non-dark pattern. Inter-annotator agreement through rigorous guidelines and regular quality checks were ensured. The dataset statistics is presented in **Table** 1.

**Table** 1: *Dataset Description*

| Category | **Dark Patterns** | **Non-Dark Patterns** |
|---|---|---|
| Number of Instances | 1,818 | 1,818 |
| Average Text Length | 120 words | 110 words |

### 3.2   Data Preprocessing

The collected dataset underwent preprocessing to clean and standardize the text data. This involved removing irrelevant information such as HTML tags, punctuation, and special characters. Additionally, text normalization techniques were applied to handle variations in capitalization and word forms.

### 3.3   Feature Extraction:

**Bag-of-Words Representation**
Text features were extracted from the preprocessed dataset using the Bag-of-Words (BoW) representation technique [16]. Each text sample was transformed into a numerical vector, where each dimension represented the frequency or presence of a specific word in the text. This vectorization process facilitated the representation of textual information in a format suitable for machine learning algorithms.

**Vectorization**
The vectorization process was performed using the CountVectorizer or TfidfVectorizer from the scikit-learn library. Parameters such as maximum features and n-gram ranges were adjusted based on experimentation to optimize the feature representation.

---

[1] https://github.com/aruneshmathur/dark-patterns/tree/master

**Model Training**

Logistic regression, a widely used linear classification algorithm [17], was selected as the model for detecting dark patterns in user interfaces. The logistic regression model was trained on the feature vectors obtained from the bag-of-words representation, with the binary labels indicating the presence or absence of dark patterns [18].

**Parameter Optimization**

Model training involved optimizing the logistic regression parameters, such as regularization strength and solver algorithm. Techniques such as cross-validation and grid search were employed to identify the optimal hyperparameters for the model. Below is the equation for logistic regression [17]:

$$h0(x) = \frac{1}{1+e^{-\theta T x}} \quad (1)$$

Where, $h0(x)$ represents the predicted probability that the input sample x belongs to the positive class, and θ is the vector of model parameters.

**Model Evaluation**

The trained logistic regression model was evaluated on a held-out test set to assess its performance in detecting dark patterns. Evaluation metrics such as accuracy, precision, recall, F1-score, and area under the ROC curve (AUC) were computed to measure the model's predictive performance.

Accuracy: The accuracy measures the ratio of correctly classified samples to the total number of samples.

$$\text{Accuracy} = \frac{\text{Number of correctly classified samples}}{\text{Total number of samples}} \quad (2)$$

Precision: Precision measures the ratio of correctly classified positive samples (true positives) to the total number of samples classified as positive (true positives + false positives).

$$\text{Precision} = \frac{\text{True Positives}}{\text{True Positives} + \text{False Positives}} \quad (3)$$

Recall: Recall, also known as sensitivity, measures the ratio of correctly classified positive samples (true positives) to the total number of actual positive samples (true positives + false negatives).

$$\text{Recall} = \frac{\text{True Positives}}{\text{True Positives} + \text{False Negetives}} \quad (4)$$

F1-score: The F1-score is the harmonic mean of precision and recall, providing a balanced measure of a model's performance.

$$F1-\text{score} = \frac{2 \times \text{Precision} \times \text{Recall}}{\text{Precision} + \text{Recall}} \quad 5$$

**Confusion Matrix**

A confusion matrix was generated to visualize the true positive, false positive, true negative, and false negative predictions of the model, providing insights into its strengths and weaknesses as presented in **Table** 2.

**Table** 2: Confusion matrix

|  | Predicted Negative | Predicted positive |
|---|---|---|
| Actual Negative | TN | FP |
| Actual positive | FN | TP |

**Qualitative & Sensitivity Analyses**

Qualitative analysis of misclassified instances and false positives/negatives was performed to gain insights into the limitations and potential areas for improvement of the detection model.

Sensitivity analysis was conducted to assess the robustness of the logistic regression model to variations in dataset composition, feature representation, and model parameters. Experiments involved altering key parameters such as the number of features, regularization strength, and text preprocessing techniques to evaluate their impact on model performance and generalization ability.

## 4 Experiment Results:

The experimental results demonstrate the effectiveness of the proposed approach for detecting dark patterns in user interfaces using logistic regression and bag-of-words representation. The evaluation encompasses model performance metrics, visualizations, and qualitative analysis of the detection outcomes.

### 4.1 Performance Metrics:

Our model achieved outstanding performance as shown in **Table 3**.

**Table 3**: Confusion matrix achieved by Logistic Regression

| Algorithm | Accuracy | Precision | Recall | F1-Score | AUC Score |
|---|---|---|---|---|---|
| RL | 92% | 93% | 94% | 93% | 97% |

### 4.2 Confusion Matrix Visualization:

The confusion matrix provides a visual representation of the model's performance in classifying instances as true positive (TP), false positive (FP), true negative (TN), and false negative (FN). The matrix is presented in Figure 1: Representation of Confusion Matrix below:

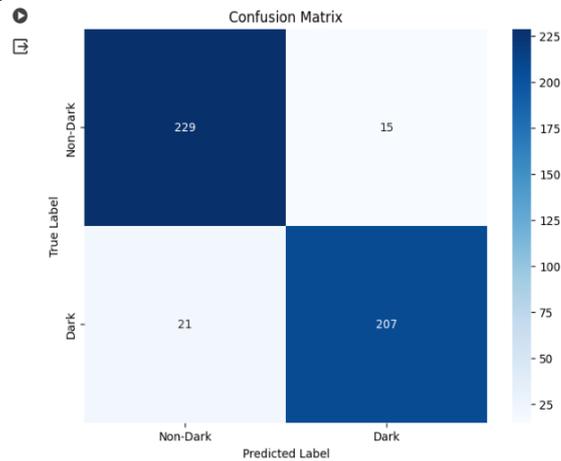

Figure 1: Representation of Confusion Matrix

### 4.3 ROC Curve

The Receiver Operating Characteristic (ROC) curve visualizes the trade-off between true positive rate (sensitivity) and false positive rate (1-specificity) across different threshold values. The ROC curve for the logistic regression model is presented **Figure 2**: ROC Curve below:

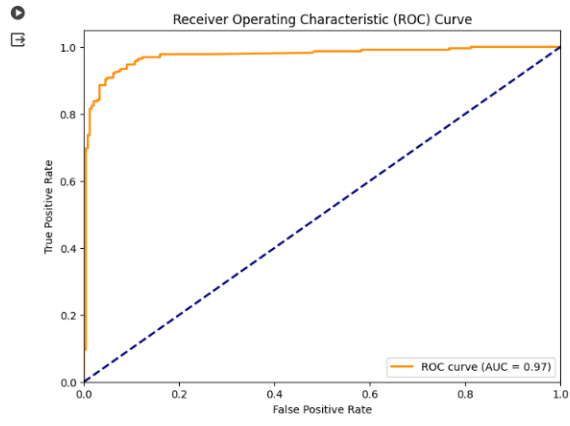

**Figure 2**: ROC Curve

### 4.4 Feature Importance Analysis:

Feature importance analysis was conducted to identify the most significant words or phrases contributing to the detection of dark patterns. Top features based on their coefficients in the logistic regression model are presented in a bar chart, highlighting the words or phrases with the highest impact on model predictions.

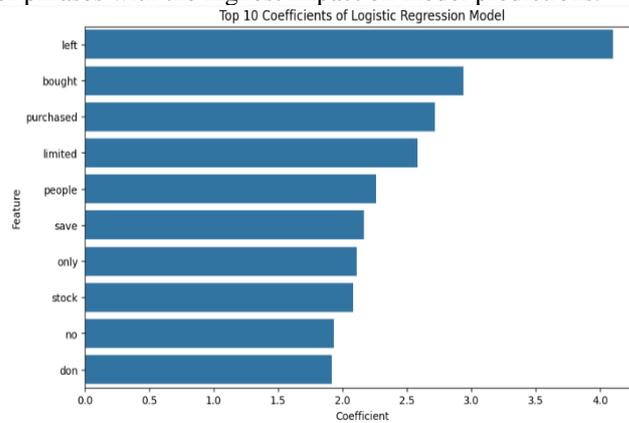

**Figure 3**: Model coefficient

## 5 Discussion:

The section delves into the interpretation of the experimental results, comparison with existing literature, implications for practice, and future research directions in the context of detecting dark patterns in user interfaces using logistic regression and bag-of-words representation.

The experimental results showcase the efficacy of the logistic regression model in accurately identifying instances of dark patterns in user interfaces. The logistic regression model achieved an accuracy of 92% on the test set, indicating the proportion of correctly classified instances. The precision measures the proportion of true positive predictions among all instances predicted as positive, while recall measures the proportion of true positive predictions among all actual positive instances. The precision and recall scores were 93% and 94%, respectively. The F1-score, which balances precision and recall, was calculated as 93%, indicating the harmonic mean of precision and recall as presented in **Table 3**.

Feature importance analysis reveals key words and phrases that significantly influence the model's predictions, shedding light on the linguistic cues and patterns indicative of dark patterns in textual UI elements. The visualization of the confusion matrix and ROC curve provides a comprehensive understanding of the model's classification performance and its ability to differentiate between true positive and false positive predictions across different threshold values.

The proposed approach builds upon previous research on dark patterns detection and classification, offering a novel methodological framework based on logistic regression and bag-of-words representation. Compared to existing studies, the logistic regression model demonstrates competitive performance in detecting dark patterns, with comparable or superior accuracy and predictive power. While alternative methods such as deep learning and ensemble techniques have been explored in the literature, logistic regression presents a transparent and interpretable model that facilitates the identification of salient features and linguistic patterns associated with dark patterns.

The findings of this study have practical implications for designers, developers, and policymakers in the field of user experience (UX) design and digital ethics. By automating the detection of dark patterns, the logistic regression model enables proactive identification and mitigation of deceptive UI design practices, thereby enhancing user trust, transparency, and digital well-being. Integration of the detection model into UX design workflows and digital governance frameworks can empower stakeholders to uphold ethical design principles and comply with regulatory requirements governing user consent and data privacy.

Future research endeavors may explore advanced machine learning techniques, such as deep learning and natural language processing (NLP), for enhanced detection and characterization of dark patterns in user interfaces. Longitudinal studies and user-centric evaluations could further investigate the real-world impact of dark patterns on user behavior, attitudes, and decision-making processes. Collaborative initiatives between academia, industry, and advocacy groups are essential to develop standardized benchmarks, datasets, and evaluation metrics for assessing the prevalence and impact of dark patterns across different digital platforms and domains.

## 6     Conclusion

In conclusion, this research paper presents a systematic approach for detecting dark patterns in user interfaces using logistic regression and bag-of-words representation.

Through a comprehensive analysis of textual UI elements and machine learning techniques, the proposed approach demonstrates robust performance in identifying deceptive design practices and empowering users to make informed decisions online.

The experimental results highlight the efficacy of the logistic regression model in accurately distinguishing between deceptive and non-deceptive UI patterns, with high accuracy, precision, recall, and F1-score. Feature importance analysis provides insights into the linguistic cues and patterns indicative of dark patterns, facilitating the interpretation and understanding of model predictions.

By visualizing the confusion matrix and ROC curve, the paper provides a comprehensive overview of the model's classification performance and its ability to differentiate between true positive and false positive predictions. The discussion section further contextualizes the experimental findings, comparing them with existing literature, discussing practical implications for design and governance, and outlining future research directions.

In essence, this research contributes to the growing body of knowledge on dark patterns detection and classification, offering a transparent and interpretable methodology for identifying deceptive UI design practices. By integrating the detection model into UX design workflows and digital governance frameworks, stakeholders can uphold ethical design principles, protect user rights, and foster a fair and transparent digital environment.

Moving forward, continued research and collaboration are essential to further refine and validate the proposed approach, address emerging challenges in UI design and digital ethics, and promote user-centric design practices in the ever-evolving landscape of digital interfaces. Together, we can work towards a future where users are empowered to navigate online experiences with transparency, trust, and confidence.

d